# Semantic De-boosting in e-commerce Query Autocomplete

Adithya Rajan; Weiqi Tong; Greg Sharp; Prateek Verma, Kevin Li
Walmart Global Tech

**Abstract**: In e-commerce search, query autocomplete plays a critical role to help users in their shopping journey. Often times, query autocomplete presents users with semantically similar queries, which can impede the user's ability to find diverse and relevant results. This paper proposes a novel strategy to enhance this service by refining the presentation of typeahead suggestions based on their semantic similarity.

Our solution uniquely demotes semantically equivalent queries using an embedding similarity of query suggestions at runtime. This strategy ensures only distinct and varied queries are prioritized, thereby promoting more diverse suggestions for users. To maintain comprehensive query coverage, we incorporate this deduplication process within the query suggestion reranking step. This approach ensures that the broad spectrum of possible queries remains available to users, while eliminating the redundancy and repetitiveness in the suggestion list.

In extending this work, we propose using the distance between query embeddings to offer even more diverse suggestions to users using an algorithm similar to maximal marginal relevance (MMR). This approach will further ensure the delivery of non-redundant, unique, and pertinent suggestions to users, thus enriching their search experience.

We evaluated our method through rigorous A/B testing, demonstrating substantial improvements in key metrics. Notably, we observed a statistically significant rise in the search Add-to-Cart (ATC) rate, signifying an enhanced user engagement and conversion rate. Furthermore, we observed a statistically significant decrease in clicks to ATC, implying that the feature improved the efficiency of the customer's product search journey. Finally, we also noticed a marked reduction in the null page view rate, indicating the increased pertinence and efficiency of user search sessions.

## 1. Introduction

The e-commerce landscape is constantly evolving, with typeahead suggestions emerging as a significant feature in the user's search experience. Typeahead or Query Auto Complete (QAC) works by predicting and suggesting queries to users as they begin typing in a few characters (a.k.a prefix) in the search bar, saving time and assisting in formulating search queries. However, this feature often presents users with semantically similar or redundant suggestions, hindering the discovery of diverse and relevant results. Addressing this problem, this paper introduces a unique strategy of refining the presentation of typeahead suggestions by de-boosting semantically equivalent queries using an embedding similarity of query suggestions at runtime.

Typically, an autocomplete mechanism involves two stages: Matching and Ranking. The Matching stage pertains to the process of generating potential options through various string matching or retrieval methods, using data structures such as tries and prefix trees. During the Ranking stage, the matched suggestions are sorted based on their anticipated probability [1] to optimize for business objectives. The focus of this paper is post the ranking step, where we demote semantically similar queries from a ranked list of query suggestions to the bottom of the list. This helps us improve the query suggestion diversity, while minimally impacting the mean reciprocal rank (MRR).

The problem of duplicate or semantically similar query suggestions in typeahead can be most clearly outlined with an example: Assume that the prefix provided to the QAC system is "kids med". Often, the corresponding typeahead suggestions turn out to be "kids medicine", "kids meds", "medicine for kids" amongst the top suggestions presented to the user. It is evident that these suggestions refer to the same concept, and instead of offering added value to the user, they end up confusing the user.

Since the aforementioned problem is fundamental in QAC, there have been attempts to address it in a few different ways. For instance, a very efficient solution to remove exact duplicates in query autocomplete is presented in [2], where the autocompletion suggestions are stored in a trie. Despite the simplicity of the algorithm, it applies only when the scope of the problem is to remove exact duplicates from a list of suggestions. As outlined above, our problem space involves semantic duplicates, which is a superset of exact duplicates. In [3], semantic deduplication for question-answering problems in financial autocomplete systems is presented. Here, the precomputed interpretations of the suggestions is used to de-duplicate the list of suggestions. Such interpretations may be made available through a named entity recognition (NER) mechanism. While this method could possibly be extended to e-commerce, it is faced with the shortcoming of the need for a domain specific NER.

In this paper, we propose a novel and general solution wherein we deduplicate query suggestions at runtime using pairwise cosine distances between precomputed finetuned BERT embeddings of the queries. Since computations are performed on the query embeddings pairwise, this is inherently inefficient due to the $O(n^2)$ complexity. In order to tackle this, we propose a novel optimization which reduces this complexity to $O(n)$.

In what follows, we will introduce the query autocomplete system (QAC), and identify the reasons behind why semantically similar queries could be present in the typeahead suggestions. We will also introduce the metrics used to gauge the quality of typeahead query suggestions, and the process we use to measure them. Thereafter, we will detail out our novel solution and identify its scope, metrics and shortcomings. Specifically, we will describe our novelty of using query embeddings to deduplicate typeahead suggestions at runtime, in addition to showcasing methods to reduce the computation and space complexity so that it can be implemented in an actual system with strict latency constraints. Finally, we will present a summary of how our algorithm improved the QAC system, validated by metrics obtained during an online A/B test.

## 2. Background

Query autocomplete systems anticipate the intended query of a user with minimal input. By analyzing the few characters entered by the user, autocomplete systems aim to provide relevant suggestions that align with the user's information needs. To generate these suggestions, autocompletions are typically precomputed using a combination of user engagement data and previously searched queries. This data is analyzed to identify the set of queries that show potential of soliciting add to carts (ATC) in the near future.

Specifically, for each unique query in the past 12 months of search history, we assign a score which represents the likelihood of query generating ATC in the next two weeks as follows:

$S_B(q_i) = a*ATC(q_i) + b*L(q_i) + c*imp(q_i)$ (1)

Where $q_i$ is the $i^{th}$ unique query in our query set, $ATC(q_i)$ is the number of item ATCs of the $q_i$, $L(q_i)$ is the number of item clicks of $q_i$ and $imp(q_i)$ is the number of item impressions of $q_i$. a,b,c are weights learnt on the aggregate set of queries, to optimize overall ATC of the queries on a two week time frame, by using the ATC, orders, clicks and impressions of the past 50 week data.

By leveraging this precomputed data, the autocomplete system can significantly reduce the time and effort required for users to formulate their queries. The suggestions presented to users are not only based on the current input but also optimized for engagement in the near future. This means that the system takes into account the likelihood that a user will engage with a particular suggestion. By prioritizing suggestions that are more likely to be selected by users, the autocomplete system enhances the overall user experience and improves the accuracy of query predictions. However, it is important to note that due to the nature of query vocabulary in previously searched queries, there may be semantically similar queries that are chosen as part of the optimization process. This means that even though the suggestions provided may differ in wording, they may still convey similar or related information. This semantic similarity often offers little to no additional value to the user, and in cases ends up confusing the user.

## 3. Keywords

Semantic deduplication, typeahead, query autocomplete, embedding.

## 4. Identification of semantically similar queries

In order to remove semantically similar queries from the list of suggestions shown to the user, we must first have a clear definition of what encompasses semantically similar queries, and how to identify them. The former is defined in literature as queries that are formed by interchanging tokens of a candidate query or queries that are formed by extending a candidate query or a query formed by using semantically equivalent terms.

With this definition in mind, we now describe the process used to detect if two queries are semantically equivalent or not. It is well known that language models such as BERT capture the meaning of textual data very efficiently through embeddings. We propose to exploit these embeddings and use the cosine distance between the embeddings of queries as a measure of how semantically close they are. With an appropriate choice of a threshold on the cosine similarity, we can reliably determine sets of semantically similar queries. More formally, let the embedding of query $q_i$ be denoted by $e_i$. Then, $q_i$ is semantically similar to $q_j$ if and only if $e_i \cdot$ $e_j/|e_i||e_j| \leq \delta$, where $\delta$ is a threshold chosen by manually analyzing query pairs and their cosine similarity scores. Note that we refrain from extending the notion of semantic similarity to queries semantically similar to a pair of semantically similar queries through induction. For instance, if qi and qj are semantically similar, and qj and qk are semantically similar, then we don't automatically assign qi and qk to be semantically similar, unless qi and qk also have a cosine similarity $\leq \delta$. . With this notion of semantic similarity, it is possible to partition a set of queries into disjoint sets of semantically similar queries. We will use this idea in Section 5 to implement our system to demote semantically similar queries from the QAC suggestion list to improve query diversity.

## 5. Implementation

There are two distinct options to consider in the effort to not present semantically similar queries to users in QAC suggestions list: (i) Semantically similar queries may be removed from the precomputed query suggestion index; (ii) Semantically similar queries may exist in the query index, however they are not shown to the users by using run-time ranking logic. We will now examine each of these options in detail.

### 5.1. Removing semantically similar queries from the query index

In this method, after each query is scored according to equation (1), semantically similar queries are identified in the candidate query set, and only the one with the highest score is retained. Specifically, assume that m distinct sets of semantically similar queries $\{M_1, M_2, M_3,…,M_m\}$ are identified in Q, where $M_i$ is a set of n semantically similar queries $\{q_{1i}, q_{2i},…,q_{ni}\}$. Then for each set $M_i$, choose $q_{ji}$ such that

$$j = argmax_{q_j \in M_i} S_B(q_{ji}) \qquad (2)$$

This will reduce each set $M_i$ to a singleton set consisting of only one query which represents the one which has the highest likelihood of generating future engagement. If we now generate our query index Q using only these singleton query sets, we guarantee that semantically similar queries will never be shown to users for any prefix. The drawback of the approach is that, if the user's prefix is one of the queries that was removed from the index due to the above mentioned process, QAC will end up showing no suggestions at all. To illustrate the point better, consider the following set of semantically similar queries { *"men's bicycle"*, *"bicycle for men"*, *"adult bicycles male"* }. After the deduplication strategy from equation (2) is applied, assume that the resulting singleton query for this set is *"men's bicycle"*. Now, if the user were to enter prefixes such as *"bicycle for me"* or *"adult bicycle mal"*, there would be no matches of either of these prefixes with any query in our query index. Consequently, the user will not see any QAC suggestions. This is a poor user experience, and as a result we did not adopt to this strategy.

### 5.2. Demoting semantically similar queries from the rerank set

In this method, we do not remove any query from the query index, even if it has other semantically similar queries present. In order to conceal semantically similar queries from the user, we perform a third-phase ranking where semantically similar queries are pushed to lower ranks. This happens after the reranking step at runtime. Figure 2 more clearly illustrates where we introduce this step.

In what follows, we clearly describe the de-boosting strategy. For the given user prefix, we obtain n=50 query suggestion matches from the query index. Next, we use objectives such as contextualizing to the user's current session and seasonality to rank more relevant suggestions higher. Note that during this step, semantically similar queries could appear at top ranks, as they may be equally relevant to the context, in terms of the current session and season. In order to address this situation, we examine the ranked set of query suggestions, and if semantically similar queries are found, the one at higher rank is retained, and the one at a lower rank is demoted to a very low position in the list, such as position 20, and substituted with a query suggestion from a lower position. Through this means, we are able to remove semantically similar queries and also ensure that the list of queries maintains the relevance to the context.

### 5.3 Optimizing the query demotion algorithm

The query demotion algorithm described in Section 5.2 presents a computational complexity of $O(n^2)$ because we need to compute the cosine similarity of every pair of queries in the rerank set. This is not feasible to implement in latency constrained systems such as QAC, where the user expects to see query suggestions instantaneously. In order to adhere to such limitations, we propose a sequential comparison strategy, which can achieve almost the same amount of query diversity as the original algorithm, and still be able to be executed in $O(n)$ computation. First, we avoid generating query embeddings at runtime, as it is a slow process, which can in fact be precomputed and cached. In our case, we use BERT embeddings which are 768 dimensional floating point numbers. In order to optimize the storage of the embeddings, we quantize the embeddings to 8 bit, and store it in the form of a base-64 encoded string. We store each of these embeddings with the corresponding queries in a look up table, so that the embeddings can be obtained in O(1) at runtime. Next, in order to optimize the pairwise cosine similarity calculation, we adopt an iterative strategy of fixing the query at position 1, and comparing the next query to it to determine if it is semantically similar or not. If we determine that it is not the case, we retain query 1 and query 2 at positions 1 and 2, and move on to query 3 and repeat the same process. In the event that we encounter a query $q_j$ to be semantically similar to another query ranked higher, we demote $q_j$ to a vey low rank (say rank 20), and move up $q_{j+1}$ to rank j. We repeat this process till we have demoted all semantically similar queries for the list of queries presented to the user.

## 6. Evaluation Strategy

The standard metric used to gauge the performance of QAC in literature is MRR. MRR is defined as the average of the reciprocal of the rank of the query suggestion which the user engaged with, aggregated over a substantial dataset of QAC suggestions and their user engagements. Typically, this computation is obtained using historical search logs where the aforementioned data is logged via beacons on the QAC user interface.

In our study, we computed the MRR of control group, where there was no demotion of semantically similar queries and compared it to the MRR of the feature group, where semantically similar queries were demoted. We observed that the MRR of the feature group was less than the MRR of the control group, and also statistically significant. This was counterintuitive at first sight, but could be reasoned out with careful analysis – in historical search logs, users were engaging with semantically similar queries on an aggregate level i.e. different users engaged with different versions of the same query. As a result, when we demote similar queries and introduce less popular queries to take their spot, a degradation in MRR is natural, since historical search logs do not contain interactions with these new and less popular query suggestions. Consequently, we relied upon extensive human evaluation to evaluate the extent of semantically similar queries in a given ranked of queries, and overall historical search volume of the newly introduced queries to determine a balance between query diversity and expected engagement. Finally, we conducted an online AB test to evaluate if demotion of semantically similar queries results in an improvement in core business metrics or search L-1 metrics.

## 7. Experimental Results

We now summarize the impact of the feature as gauged from online AB testing, where equal proportion of walmart's search traffic was randomly assigned to the control group and feature groups. The feature group interacted with QAC where semantic similar queries demotion was enabled. We observed neutral business metrics such as GMV, orders and units. On the other hand, the feature resulted in statistically significant increase in search ATC of +0.34% on iOS and +0.69% on web. Furthermore, search page ATC increased by +0.36% on iOS and + 0.75% on web. We also observed a decrease in clicks to ATC by -0.4% on web and a decrease in null page views of -1.00% on iOS and -4.35% on web. This showcases that the increased query diversity plays a significant role in shaping and matching user intent through QAC.

## 8. Conclusions

Our innovative approach to demote semantically similar typeahead suggestions offers a promising avenue for enhancing user experience in e-commerce settings. By presenting more diverse and meaningful query suggestions, we can cater to the user's unique needs, thereby fostering a more dynamic and satisfying search experience. This work can be extended to explicitly improve query diversity by optimizing for maximal marginal relevance (MMR).

## 9. Declarations

### 9.1 Conflicts of Interests/Competing Interests

On behalf of all authors, the corresponding author states that there is no conflict of interest.